\documentclass[12pt,preprint]{aastex}

\def\ltsima{$\; \buildrel < \over \sim \;$}
\def\gtsima{$\; \buildrel > \over \sim \;$}
\def\lsim{\lower.5ex\hbox{\ltsima}}
\def\gsim{\lower.5ex\hbox{\gtsima}}
\def\lapp{\ifmmode\stackrel{<}{_{\sim}}\else$\stackrel{<}{_{\sim}}$\fi}
\def\gapp{\ifmmode\stackrel{>}{_{\sim}}\else$\stackrel{<}{_{\sim}}$\fi}

\usepackage{amsmath}
\usepackage{textcomp}
\newbox\grsign \setbox\grsign=\hbox{$>$} \newdimen\grdimen \grdimen=\ht\grsign
\newbox\simlessbox \newbox\simgreatbox
\setbox\simgreatbox=\hbox{\raise.5ex\hbox{$>$}\llap
     {\lower.5ex\hbox{$\sim$}}}\ht1=\grdimen\dp1=0pt
\setbox\simlessbox=\hbox{\raise.5ex\hbox{$<$}\llap
     {\lower.5ex\hbox{$\sim$}}}\ht2=\grdimen\dp2=0pt

\newbox\simppropto
\setbox\simppropto=\hbox{\raise.5ex\hbox{$\sim$}\llap
     {\lower.5ex\hbox{$\propto$}}}\ht2=\grdimen\dp2=0pt

\begin{document} 
\title{High resolution reddening map in the direction of the stellar system Terzan 5 \footnote{Based on observations
with the NASA/ESA {\it HST}, obtained at the Space Telescope Science Institute, which operated by AURA, Inc., under
NASA contract NAS5-26555.}}

\author{
Davide Massari\altaffilmark{1},
Alessio Mucciarelli\altaffilmark{1},
Emanuele Dalessandro\altaffilmark{1},
Francesco R. Ferraro\altaffilmark{1},
Livia Origlia\altaffilmark{2},
Barbara Lanzoni\altaffilmark{1},
Giacomo Beccari\altaffilmark{3},
R. Michael Rich\altaffilmark{4},
Elena Valenti\altaffilmark{3},
Scott M. Ransom\altaffilmark{5}
}

\affil{\altaffilmark{1}Dipartimento di Astronomia, Universit\`a degli
  Studi di Bologna, via Ranzani 1, I-40127, Bologna, Italy}
\affil{\altaffilmark{2}INAF-Osservatorio Astronomico di Bologna, via
  Ranzani 1, 40127, Bologna, Italy} 
\affil{\altaffilmark{3}European Southern Observatory,
  Karl-Schwarzschild-Strasse 2, 85748 Garching bei M\"{u}nchen,
  Germany} 
\affil{\altaffilmark{4}Department of Physics and Astronomy,
  PAB 430 Portola Plaza Box 951547, UCLA, Los Angeles, CA 90095-1562, USA}
\affil{\altaffilmark{5}National Radio Astronomy Observatory,
  Charlottesville, VA 22903, USA}


\date{}

\begin{abstract}
We have used optical images acquired with the Hubble Space Telescope
to construct the first high-resolution extinction map in the direction
of Terzan 5, a peculiar stellar system in the inner bulge of our
Galaxy. The map has a spatial resolution of $8\arcsec \times
8\arcsec$, over a total FoV of $200\arcsec\times
200\arcsec$. The absorption clouds show a patchy structure on a
typical scale of $20\arcsec$ and extinction variations as large as
$\delta E(B-V)\sim 0.67$ mag, especially in the direction of the center of
the system.  These correspond to an absolute color
excess ranging from $E(B-V)=2.15$ mag, up to 2.82 mag.  After the correction
for differential reddening, two distinct red giant branches become
clearly visible in the color magnitude diagram of Terzan 5 and they
well correspond to the two sub-populations with different iron
abundances recently discovered in this system.

\end{abstract} 


\section{Introduction} 
\label{intro}
Terzan 5 is a stellar system commonly catalogued as a globular cluster
(GC), located in the inner bulge of our Galaxy 
(its Galactic coordinates are $l=3.8395$, $b=1.6868)$, at a distance of 5.9 Kpc \citep{valenti}. 
It is affected by
severe differential reddening \citep{ortolani96}, with an
average color excess $E(B-V)=2.38$ \citep{barbuy, valenti}.
\citet[][hereafter F09]{f09} discovered the presence of two distinct sub-populations,
which define two red clumps (RCs) clearly separated in luminosity in the
$(K,J-K)$ color-magnitude diagram (CMD) and show significantly
different iron content: the bright RC at $K = 12.85$ is
populated by a quite metal rich (MR) component ([Fe/H]$\simeq +0.3$),
while the faint clump at $K = 13.15$ corresponds to a relatively metal
poor (MP) population at [Fe/H]$\simeq -0.2$. Before this discovery,
such a large difference in the iron content ($\Delta$ [Fe/H]$>0.5$
dex) was found only in $\omega$ Centauri, a GC-like system in
the Galactic halo, now believed
to be the remnant of a dwarf galaxy accreted by the Milky Way.
\cite{origlia} presented a detailed study of the abundance patterns of
Terzan 5, demonstrating that (1) the abundances of light elements
(like O, Mg, and Al) measured in both the sub-populations do not
follow the typical anti-correlations observed in genuine GCs; (2) the
overall iron abundance and the $\alpha-$enhancement of the MP
component demonstrate that it formed from a gas mainly enriched by
Type II supernovae (SNII) on a short timescale, while the progenitor
gas of the MR component was further polluted by SNIa on longer
timescales; (3) these chemical patterns are strikingly similar to
those measured in the bulge field stars.

These observational results demonstrate that Terzan 5 is not a genuine
GC, but a stellar system that has experienced complex star formation and
chemical enrichment histories. Indeed it is likely to have been much more
massive in the past than today (with a mass of at least a few
$10^7-10^8 M_\odot$, while its current value is $\sim 10^6
M_\odot$; \citealt[][hereafter L10]{l10}), thus to retain the high-velocity gas ejected
by violent SN explosions. Moreover the collected evidence indicates
that it formed and evolved in strict connection with its present-day
environment (the bulge)\footnote{The probability that Terzan 5 was
  accreted from outside the Milky Way (as supposed for $\omega$
  Centauri) is therefore quite low.}, thus suggesting the possibility
that it is the relic of one of the pristine fragments that contributed
to form the Galactic bulge itself.  In this context, also the
extraordinary population of millisecond pulsars (MSPs) observed in
Terzan 5\footnote{Its 34 MSPs amount to $\sim25\%$ of the
  entire sample of MSPs known to date in Galactic GCs \citep[][; see
    the updated list at {\tt
      www.naic.edu/$\sim$pfreire/GCpsr.html}]{ransom}.} can find a
natural explanation. In fact, the large number of SNII required to
account for the observed abundance patterns would be expected to have produced
a large population of neutron stars, mostly retained by the deep
potential well of the massive {\it proto}-Terzan 5.  The large
collisional rate of this system (Verbunt \& Hut, L10) may also have favored
the formation of binary systems containing neutron stars and promoted
the re-cycling process responsible for the production of the large MSP
population now observed in Terzan 5. 

Within this exciting scenario, we are now coordinating a project aimed
at reconstructing the origin and the evolutionary history of Terzan 5.
However, severe limitations to the detailed analysis of the
evolutionary sequences in the optical CMDs are introduced by the
presence of large differential reddening. To face this problem here we
build the highest-resolution extinction map ever constructed in the
direction of Terzan 5.

\section{Differential reddening correction} 
\label{obsdata}
 
\subsection{The data-set}
The photometric data used in this work consist of a set of
high-resolution images obtained with the Wide Field Channel (WFC) of
the Advanced Camera for Survey (ACS) on board the Hubble Space
Telescope (GO-9799, see F09 and L10).  
The ACS-WFC camera has a field of
  view (FoV) of $\sim200\arcsec \times 200\arcsec$ with a plate-scale of
  0.05\arcsec/pixel. Both F606W (hereafter $V$) and F814W ($I$)
  magnitudes are available for a sample of about 127,000 stars. The
  magnitudes were calibrated on the VEGAMAG photometric system by
  using the prescriptions and zero points by \cite{sirianni}.  The
  final catalog was placed onto the Two Micron All Sky Survey
  absolute astrometric system by following the standard procedure
  discussed in previous works (e.g., L10).  The $(I,V-I)$ CMD
  shown in Figure \ref{red} clearly demonstrates the difficulty of
  studying the evolutionary sequences in the optical plane,
  because of the broadening and distortion induced by differential
  reddening. In particular, the red
  giant branch (RGB) is anomalously wide ($\Delta (V-I)\sim0.8$ mag)
  and the two RCs appear highly stretched along the reddening
  vector.

\subsection{The method}
The method here adopted to compute the differential reddening within
the ACS FoV is similar to those already used in the
literature \citep[see e.g.,][] {mcz, nataf}. Briefly, the amount of
reddening is evaluated from the shift along the reddening vector
needed to match a given (reddened) evolutionary sequence to the
reference one, which is selected as the least affected by the
extinction.  Thus, the first step of this procedure is to define the
reddening vector in the considered CMD.  It is well known that the
extinction $A_\lambda$ varies as a function of the wavelength
$\lambda$, and the shape of the extinction curve is commonly described
by the parameter $R_\lambda=A_\lambda/E(B-V)$. In order to determine
the value of $R_\lambda$ at the reference wavelengths of the F606W and
F814W filters ($\lambda_V=595.8$ and $\lambda_I=808.7$ nm,
respectively; see {\tt http://etc.stsci.edu/etcstatic/users\_guide}),
we adopted the equations 1, 3a and 3b of \citet{cardelli}, obtaining $R_V=2.83$ and
$R_I=1.82$.  With these values we then computed the reddening vector
shown in Figure \ref{red}. A close inspection
of the CMD shows that the direction
of the distortions along the RCs and the RGB is well aligned with the
reddening vector.
  
As second step, the ACS FoV has been divided into a regular
grid of $m\times n$ cells.  The cell size has been chosen small enough
to provide the highest possible spatial resolution, while guaranteeing
the sampling of a sufficient number of stars to properly define the
evolutionary sequences in the CMD. In order to maximize the number of
stars sampled in each cell, we used the Main Sequence
(MS).  After several experiments varying the cell size, we defined a
grid of 25\texttimes25 cells, corresponding to a resolution
of $8.0\arcsec\times 8.0\arcsec$. In order to minimize spurious
effects due to photometric errors and to avoid non-member stars, we considered 
only stars brighter than $V=26.6$ and with $2.7<(V-I)<3.7$ colors.  
We also set
the upper edge of the CMD selection box as the line running parallel to
the reddening vector (see Figure \ref{method}). With these
prescriptions the number of stars typically sampled in each cell is
larger than 60, even at large distance from the cluster center.

The accurate inspection of the MS population in each cell allowed us
to identify the one with the lowest extinction (i.e. where the MS
population shows the bluest average color): it is located in the
South-East region of the cluster at a distance $r\simeq80\arcsec$ from
its center. The stars in this
cell are shown in the left panel of Figure \ref{method} and those enclosed
in the selection box have been used as reference sequence for
evaluating the differential reddening in each cell.  As a "guide line"
of this sequence we used an isochrone of 12 Gyr and metallicity $Z=0.01$
\citep[from][]{marigo,girardi} suitably shifted to best-fit the MS
star distribution (see the heavy white line in Figure \ref{method}).

For each cell of the grid we determined the mean $\langle
V-I\rangle$ color and $\langle V\rangle$ magnitude. A
sigma-clipping rejection at 2-$\sigma$ has been adopted to minimize
the contribution of Galactic disc stars (typically much
bluer than those of Terzan~5) and any other interloper. Each cell is
then described by the ($\langle V-I\rangle, \langle V\rangle$)
color-magnitude pair, which defines the {\it
  equivalent cell-point} in the CMD (as an example, see the cross
marked in the right panel of Figure \ref{method}).  The relative color
excess of each $i$-th cell, $\delta[E(V-I)]_i$, is estimated by
quantifying the shift needed to move the {\it equivalent cell-point}
onto the reference sequence along the reddening vector (see
the right panel of Figure \ref{method}). From the value of
$\delta[E(V-I)]_i$, the corresponding $\delta[E(B-V)]_i$ is easily
computed using the relation
 \begin{equation}
 \delta [E(B-V)]_i=\frac{\delta [E(V-I)]_i}{(R_V-R_I)},
\end{equation}
where $i=1,m\times n$ and $m\times n=625$ is the total number of cells in
our grid.  The $V$ and $I$ magnitudes of all stars
in the $i$-th cell are then corrected by using the derived
$\delta[E(B-V)]_i$ and a new CMD is built.  The whole procedure is
iteratively repeated and a residual $\delta [E(B-V)]_i$ is calculated
after each iteration. The process stops when the difference in the color
excess between two subsequent steps becomes negligible
($\lesssim0.02$ mag). The final value of the relative color excess in each
cell $\delta [E(B-V)]_i$ is thus given by the sum over all the
iterative steps.  For robustness, we applied this
procedure in both the $(I,V-I)$ and $(V,V-I)$ planes.  The
difference between the two estimates turned out to be always smaller
than $\sim0.01$ mag and the average of the two measures was then
adopted as the final estimate of the differential reddening in each
cell.

\subsection{Error estimate and caveats}
\label{error}

Our estimate of the error associated to the color-excess in each cell
is based on the method described by {\citealp{vonbraun} (see also
  \citealp{alonso}).  We considered the uncertainty on the mean color
  of the $i$-th cell as the main source of error on the value of
  $\delta [E(B-V)]_i$. This latter was then computed as the ratio
  between the 1-$\sigma$ dispersion of the mean color and the
  parameter $a=\cos(180-\theta)$, where $\theta$ is the angle between
  the reddening vector and the color-axis.  Geometrically,
  this is equivalent to measure the difference between the values of
  $\delta [E(B-V)]_i$ of the first and last contact-points
  of the color error-bar when moved along the reddening vector
  to match the reference line.  We did not consider the
  error on the mean magnitude because, since the reference line is
  almost vertical, its contribution is negligible.  Following these
  prescriptions we obtain a typical formal error of about $0.03$ mag
  on each color excess value $\delta [E(B-V)]_i$.


A potential problem with this procedure to quantify the
differential reddening of Terzan 5 is the presence of
two stellar populations with distinct iron abundances.  Indeed, the MR
population is expected to be systematically redder than the MP one in
the CMD, and we therefore expect that at least a fraction of stars
with redder colors along the MS are genuine MR objects, and not MP
stars affected by a larger extinction.  However, by using the
\cite{girardi} isochrones, the expected intrinsic difference in the
$(V-I)$ color between the MR and MP populations is only $\delta
(V-I)\sim0.05$ mag.  Moreover, the MR population has been found to be
more centrally segregated than the MP one (F09, L10). Hence, we
expect the former to become progressively negligible with increasing
radial distance from the cluster center. On the other hand, the
uncertainties due to the photometric errors are dominant in the
central region of the system, where the two populations are comparable
in number.  Finally, the use of average values for the color and
magnitude in each cell ($\langle V-I\rangle$ and $\langle V\rangle$),
with the addition of a sigma-clipping rejection algorithm, should
reduce the effect of contamination by MR stars.  Thus, an overall
error of $0.05$ mag on the color excesses $\delta [E(B-V)]_i$ is
conservatively adopted to take into account any possible residual
effects due to the presence of a double population in Terzan 5.

\section{Results}
\label{results}
 
The final differential reddening map in the direction of Terzan 5 is
shown in Figure \ref{map}, with lighter colors indicating less
obscured regions and the center of gravity and core radius (L10)
also marked for reference.  We find that, within the area covered by
the ACS-WFC, the color excess variations can be as large as $\delta
E(B-V)=0.67$ mag.  This is consistent with the value of 0.69 mag estimated
by \cite{ortolani96} from the elongation of the RC.  The
obscuring clouds appear to be structured in two main dusty patches:
the first one is located in the North-Western corner of the map at
$30\arcsec-35\arcsec$ from the center, with an average
differential extinction $\delta E(B-V)>0.4$ mag and a peak value of
0.67 mag.  The second one is placed in the South-Eastern corner, with
typical values of $\delta E(B-V)\sim0.3$ mag.  These two
regions seem to be connected by a bridge-like structure with
$\delta E(B-V)\gtrsim 0.2-0.3$ mag.

We used this map to correct our photometric catalogue. Figure
\ref{cmd} shows the comparison between the observed (left panel) and
the differential-reddening corrected (right panel) CMDs in the
$(V,V-I)$ plane. After the correction, both the color extension of the RC and the RGB width are
significantly reduced by 40\% and $>50$\%, respectively, and V magnitudes
become $\sim0.5$ mag brighter. 
To properly quantify the effect of such a correction on
the MS width, we selected the stars along an almost vertical portion
of MS and compared their color distributions before and after the
correction.  To this end, we selected stars with $25<V<25.5$ in the
observed CMD, and 0.5 mag brighter in the
corrected one (see the dashed lines in Figure \ref{cmd}).  The
result is shown in the bottom panels of the figure. Before the
correction the MS color distribution is well represented by a Gaussian
with a dispersion $\sigma=0.18$, significantly larger than the
photometric error at this magnitude level ($\sigma_{phot}\sim0.13$).
Instead, the intrinsic width of the corrected MS is well reproduced by
the convolution of two Gaussian functions separated by 0.05 mag in
color, with a ratio of 1.6 between their amplitudes, and each one
having $\sigma=0.13$ equal to the photometric error. Such
a color separation corresponds to what expected for two stellar
populations with metallicities equal to those measured in Terzan 5
(see Sect. \ref{error}).  The adopted ratio between the amplitudes
corresponds to the number counts ratio between MP and MR populations
(L10). Hence, these two
Gaussian functions correspond to the two sub-populations at different
metallicities observed in Terzan 5.  Note that the corrected MS color
distribution shows an asymmetry
toward the redder side, which is more pronounced in
the centre of the system and decreases at progressively
larger distances. The highest amplitude Gaussian (corresponding to the
MP population) is unable to properly account for this feature, while
the convolution with the reddest and lowest amplitude Gaussian
(corresponding to the MR population, which is observed to decrease in
number with increasing distance from the centre) provides an excellent
match.

The derived reddening correction was also applied to the $(K,V-K)$ CMD
obtained from the combination of the ACS and
near-infrared data (see F09). Figure \ref{rgbs} shows the corrected CMD 
with two well separated RGB sub-populations and the two
distinct RCs.  The ratio between the number of stars counted
along the two RGBs is $\sim1.5$, in very good agreement
with the value from the RCs (see above and L10).

The differential reddening corrected CMD can be finally used to
estimate the absolute color excess in the direction of Terzan
5.\footnote{A free tool providing the color excess values at any
  coordinate within the ACS-WFC FoV can be found at the web site {\tt
    http://www.cosmic-lab.eu/Cosmic-Lab/}} Different values of
$E(B-V)$ are provided in the literature, ranging from 1.65 (estimated by
\citealt{az} from the strength of an interstellar band at 8621\AA), up
to 2.19 and 2.39, derived from optical or infrared photometry
\citep[][]{barbuy,cohn,valenti}. However, all these estimates are
average values and do not take into account the presence of
differential reddening.  Here, instead, we want to build a
2-dimensional map of the absolute reddening and, to this end, we
shifted the corrected $(V,V-I)$ CMD of Terzan 5 along the reddening
direction until it matches the CMD of 47 Tucanae, adopted as reference
cluster since it is metal-rich, low extincted and with a well-determined distance modulus.  
In particular we
looked for the best match between the RC of the MP population
of Terzan 5 and the RC of 47 Tucanae.  We adopted the color
excess $E(B-V)=0.04$ and the distance modulus $\mu_0=13.32$ for 47
Tucanae \citep[from][]{ferr99}, and $\mu_0=13.87$ for Terzan 5
\citep{valenti}.
From \citet{girardi} model, in the $(V,V-I)$
plane the RC of 47 Tucanae turns out to be 0.02 mag brighter and 0.03 mag bluer
than the MP one of Terzan 5 because of a difference in their metallicity
([Fe/H]$=-0.70$ for 47 Tucanae and 
[Fe/H]$=-0.27$ for the MP population of Terzan 5; see \citealt{ferr99} and
\citealt{origlia}, respectively).
Taking into account these slight differences, a
nice match of the two RCs is obtained by
adopting $E(B-V)=2.15$ mag. Since the corrected CMD is, by construction,
referred to the bluest cell, the absolute color excess within the
ACS-WFC FoV varies from $E(B-V)=2.15$ up to
$E(B-V)=2.82$ mag. 
In order to check the reliability of these estimates, we compared it with
the values found by \citet{gonz} from the Vista Variable in the Via Lactea
survey. In a $2\arcmin \times 2\arcmin$ region centered on Terzan 5, 
these authors found an extinction $A_{K}=0.80$ mag.
Using \citet{cardelli} coefficients to convert $E(B-V)$ to $A_{K}$, our estimate varies from
$A_{K}=0.75$ to $A_{K}=0.98$ mag, in nice agreement with \citet{gonz} result.

Moreover, we looked for a possible correlation between the color excess and the 
dispersion measures for 34 MSPs of Terzan 5 studied by \cite{dm}. 
In this case
we did not find a strong correlation, probably because mostly 
(75\%) of the MSP sample is situated within the inner $20\arcsec$ of the system, where the
estimate of $E(B-V)$ is more uncertain (see Sect. \ref{error}).
 
\acknowledgements{ We thank the anonymous referee for his/her useful comments. 
This research is part of the project COSMIC-LAB
  funded by the European Research Council (under contract
  ERC-2010-AdG-267675). R.M.R aknowledges support from the National
Science Foundation from grant AST-0709479.
This publication uses data from the Two Micron All Sky
Survey, a project of the University of
Massachusetts and the Infrared Processing and Analysis Center/California
Institute of Technology, founded by the National Aeronautics and Space
Administration and the National Science Foundation.
}
 
\newpage
 \begin{figure}[!htp]
  \centering
  \includegraphics[scale=0.5]{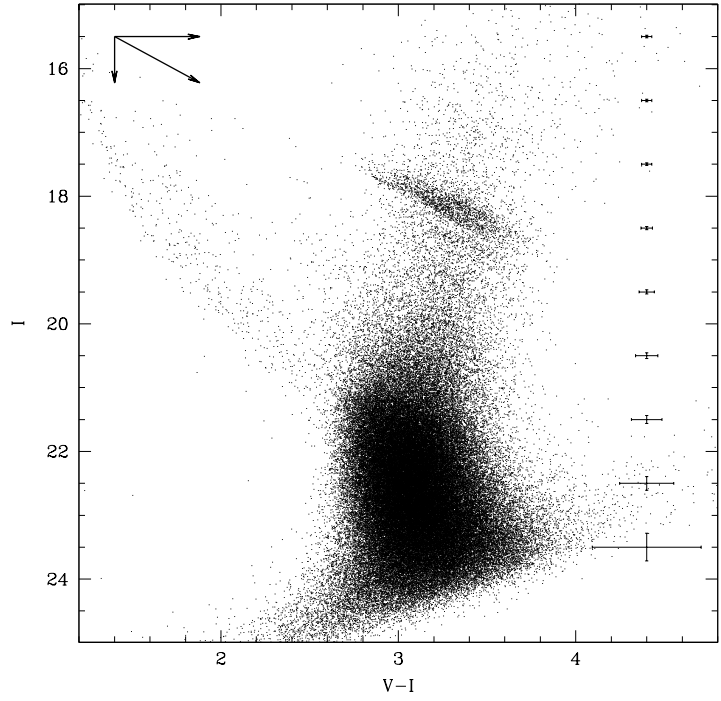}
  \caption{\small $(I,V-I)$ CMD of the $\sim127,000$ stars measured
    in the ACS FoV. The photometric errors at different magnitude levels 
    are shown.  
    Note how the distortions of the
    evolutionary sequences (in particular the RCs) 
    follow the reddening vector, shown in the
    upper left corner.}
\label{red}
\end{figure}


\newpage
 \begin{figure}[!htp]
  \centering
  \includegraphics[scale=0.5]{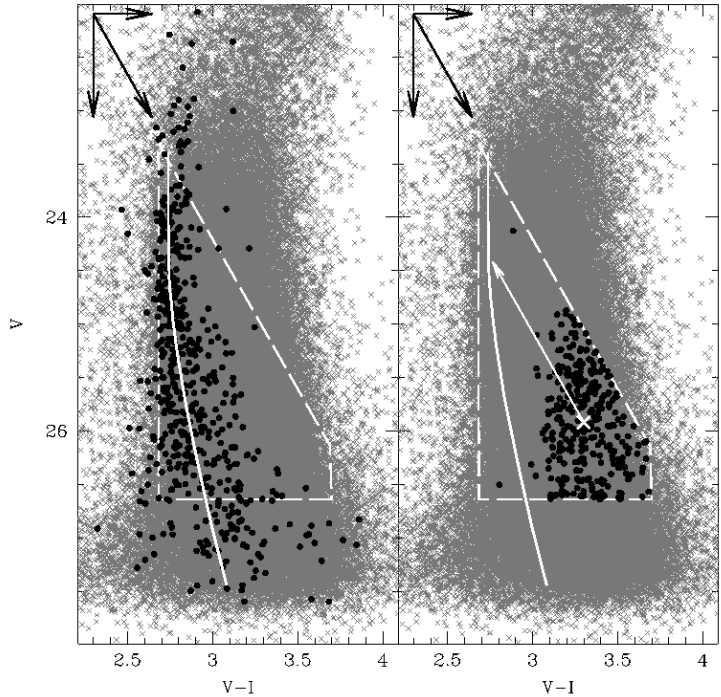}
  \caption{$(V,V-I)$ CMD of Terzan 5 zoomed in the MS region (grey
    crosses). The white dashed lines (same in both panels) delimit the
    selection box for the stars used for the
    computation of the differential reddening correction in each cell.
    The reddening vector is shown in the upper left
    corners of the diagrams. {\it Left panel--} The stars in
    the least extincted (bluest) cell are marked as
    black dots and their best-fit is shown as
    a white solid line.  {\it Right panel--} The mean color and
    magnitude of the stars selected in the $i$-th cell (black dots)
    define the {\it equivalent cell-point} ($\langle V-I\rangle$,
    $\langle V \rangle$), marked as a white cross.  The color
    excess of the cell is obtained by quantifying the shift
    needed to project this point onto the reference line along the
    reddening vector.}
\label{method}
\end{figure}

\newpage
\begin{figure}[!htb]
 \centering
 \includegraphics[height=13cm,width=15cm]{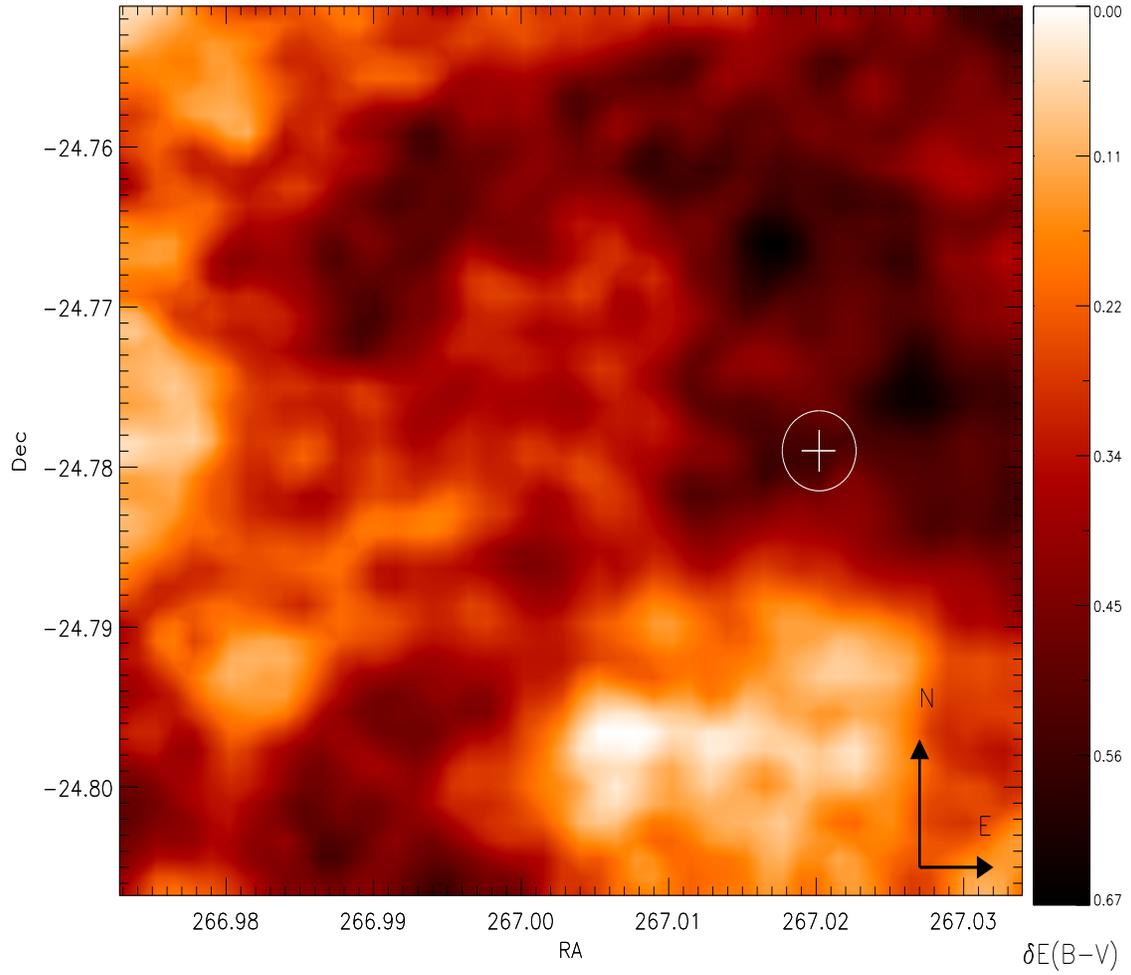}
 \caption{\small Reddening map of the ACS-WFC FoV
   ($200\arcsec\times200\arcsec$) in the direction of Terzan 5. The
   differential color excess $\delta E(B-V)$ ranges between zero (lightest)
   and 0.67 (darkest).  The gravity center and core
   radius of Terzan 5 (L10) are marked for reference as
   white cross and circle, respectively.}
\label{map}
\end{figure}

\newpage
\begin{figure}[!htb]
 \centering
 \includegraphics[scale=0.5]{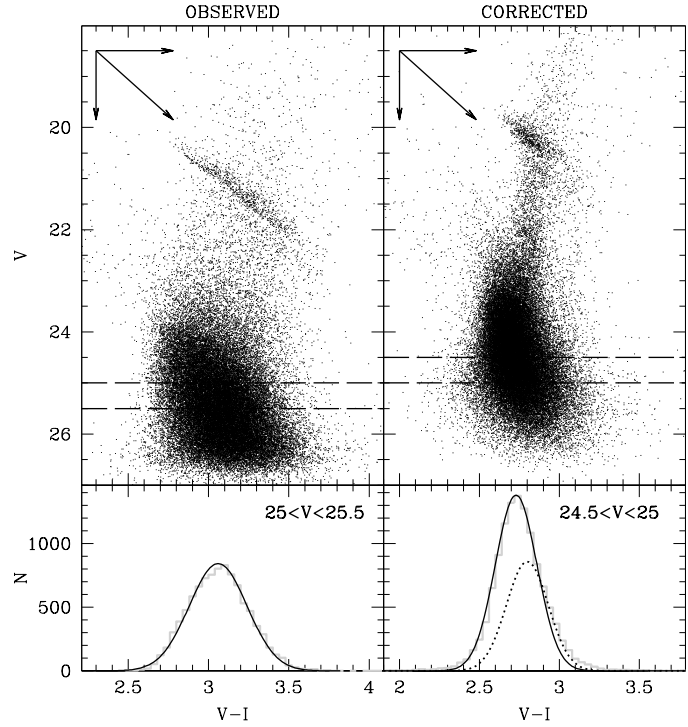}
 \caption{\small Comparison between the optical CMDs of Terzan 5
   before (left panel) and after (right panel) the differential
   reddening correction.  Only stars located at a distance
   $20\arcsec<r<80\arcsec$ are plotted for sake of clarity. All the
   sequences in the corrected CMD are much less stretched along the
   reddening vector.  The bottom panels show the color
   distributions (grey histograms) for a nearly vertical portion of
   MS at $25<V<25.5$ in the observed CMD, and at
   $24.5<V<25$ in the corrected one (see the
   dashed lines in the two upper panels).  Before the correction, the
   color distribution is well represented by a Gaussian with
   $\sigma=0.18$ (while the photometric error is
   $\sigma_{phot}\sim0.13$). After the correction, the distribution is well fitted
   by the convolution of two
   Gaussian functions with $\sigma=0.13$, separated by 0.05 mag in color and with an amplitude ratio of
   1.6. The solid Gaussian
   correponds to the MP population of Terzan 5,
   while the dotted one represents the MR component (Sect. \ref{results}).}
\label{cmd}
\end{figure}

 
\newpage
\begin{figure}[!htb]
 \centering%
 \includegraphics[scale=0.5]{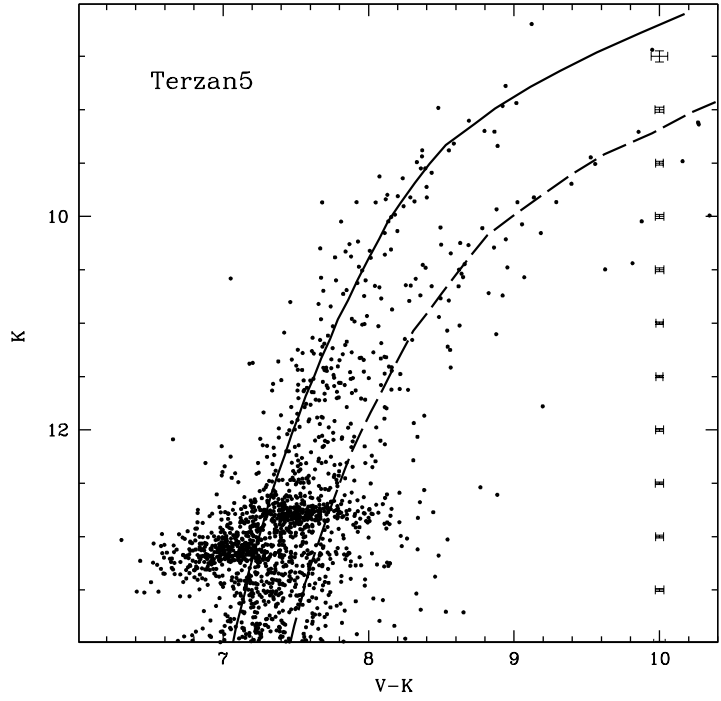}
 \caption{\small Brightest portion of the differential reddening corrected
  $(K,V-K)$ CMD of Terzan 5, with error bars also reported.  
   Beside the two RCs, also two well separated RGBs are clearly
   distinguishable. The solid and dashed lines correspond to the mean
   ridge lines of the MP and the MR sub-populations, respectively. }
\label{rgbs}
\end{figure}

{}

\end{document}